\documentstyle[psfig]{l-aa}

\def\apgt{\ {\raise-.5ex\hbox{$\buildrel>\over\sim$}}\ }
\def\aplt{\ {\raise-.5ex\hbox{$\buildrel<\over\sim$}}\ }

\newcommand{\msun}{\mbox{${\rm M}_\odot$}}

\newcommand{\rsun}{\mbox{${\rm R}_\odot$}}
\newcommand{\kms}{\mbox{${\rm km~s}^{-1}$}}

\def\unit#1{{\mbox{[{\rm #1}]}}}

\def\unit#1{{\mbox{[{\rm #1}]}}}

\begin{document}

\thesaurus{03.13.6;
	   04.03.1;
	   08.02.3;
	   08.16.6;
	   08.16.7}

\title{Observational Constraints on Kicks in Supernov\ae}
\author{Simon~F.~Portegies~Zwart \inst{1, 2},
Marco~L.~A.\ Kouwenhoven \inst{2} \& Alastair~P.\ Reynolds\inst{2, 3}}
\offprints {Simon~Portegies~Zwart: spz@astro.uva.nl}
\institute{Astronomical Institute {\em Anton Pannekoek}, 
       Kruislaan 403, 1098 SJ Amsterdam, The Netherlands
\and
	 Astronomical Institute, Postbus 80000,
                3508 TA Utrecht, The Netherlands 
\and
	Astrophysics Division, Space Science Department of ESA, ESTEC,
	Keplerlaan 1, 2200 AG Noordwijk, The Netherlands 
}

\date{Received; accepted:}

\maketitle
\markboth{Simon~Portegies~Zwart et al.\, :Constraints on Kicks}{}

\begin{abstract}
The absence or presence of extremely wide binaries with a radio pulsar
and an optical counterpart imposes a strong constraint on the existence
and magnitude of kicks in supernova explosions. We search for such
systems by comparing the positions of radio pulsars which are not
known to be in binaries with the positions of visible stars, and find
that the number of associations is negligible. 
According to the performed population synthesis, this implies that kicks
must occur, with a lower limit of at least 10 to 20 km\,s$^{-1}$.

The single 13-th magnitude star at a distance of 4.9 seconds of arc
from the pulsar PSR~B1929+10 is a good candidate to be the member of
such a wide pair.  If it turns out that this pulsar is
indeed the member of a wide binary or if another wide pair will be
found in the future, the kick-velocity distribution must have a
significant contribution from low-velocity kicks.

\keywords{Methods: statistical -- 
	  Catalogs -- 
	  Binaries: general --
	  Pulsars: general -- 
	  Pulsars: individual: PSR~B1929+10}

\end{abstract}

\section{Introduction}
Neutron stars are believed to receive high velocities upon their
formation in a supernova explosion (Gunn \& Ostriker
1970). These ``kicks'' can have a dramatic influence on
the evolution of high-mass binaries, preserving binaries which would
otherwise be disrupted in the supernova, or splitting binaries which
would otherwise remain bound.

There is no direct evidence that kicks happen: they are
inferred from e.g.\ studies of the proper motion of single pulsars
(Lyne \& Lorimer 1994), fitting observed pulsar characteristics
(Hartman 1997), or by explaining a precessing pulsar orbit
(Kaspi 1996, see van den Heuvel \& van Paradijs 1997, for a brief
overview). 
According to Iben \& Tutukow (1996) there are no kicks at all
and pulsars are only formed in type Ib and Ic supernov\ae\ from close
interacting binaries see, however, Portegies Zwart \& van den Heuvel
(1997) for counter arguments. 

Our method hinges on the fact that wide binaries are very fragile and
therefore sensitive to kicks.  Such systems, which are sufficiently
detached not to experience a phase of mass transfer, might very well
survive the first (and possibly also the second) supernova explosion
due to e.g.\ the primordial eccentricity of the binary, provided kicks
are absent (Portegies Zwart \& Verbunt 1996).  In that case it
consists of a radio pulsar in a wide binary with a main-sequence star.
Since the components will be widely separated and have small relative
orbital velocities, they will be easily mistaken for single stars.
Once kicks are considered, however, the survival probability of the
binaries is sharply reduced since the relative orbital velocity of
very wide pairs is small ($v_{\rm orb} \la 1\kms$) compared to the
expected average value of kick velocities.  Hence, the existence of
wide binaries can in principle constrain the validity of kick models.
Such systems, if they exist at all, could be identified by searching
for positional correlations between radio pulsars and ``single''
stars. By comparing the expected frequency of such systems
with the number identified in catalog comparisons we can
ascertain whether kicks play a role, and, if so, place constraints on
the lower limit of their velocities.

In the next section we discuss the fraction of such binaries that is
expected among radio pulsars.  In section~3 we search for positional
coincidences between radio pulsars and ``single'' stars using the
Taylor et al.\ (1993, 1995) catalog and the Hubble Space Telescope
Guide Star Catalog (Lasker et al.\ 1988 and Jenkner et al.\ 1990).
Finally the results are 
discussed in section~4 and we derive a lower limit to the occurrence
of kick velocities.

\section{The survival of wide binaries}
In a binary which is too wide for any Roche-lobe overflow to occur the
primary will first explode in a type~II or type Ib supernova.  A
primary with a mass between 8~\msun\ and 40~\msun\ (van den Heuvel \&
Habets 1984) is expected to leave behind a radio pulsar. If the
binary is eccentric and the eccentricity is conserved until the
supernova, the binary has a fair chance to survive and stay bound as a
binary with a main-sequence star and a young radio pulsar. However, if
there are kicks then the survival probability becomes much smaller.
We attempt to model the surviving fraction of wide binaries using the
population synthesis model of Portegies Zwart \& Verbunt (1996, see
also Portegies Zwart \& Yungelson 1997)

We assumed the following initial conditions (Abt 1983; Duquennoy \&
Mayor 1991): the primary mass is chosen from a power-law
distribution with an exponent $\alpha = -2.5$ (Salpeter is -2.35)
between 8 and 100~\msun.
The semi-major axis distribution was taken to be uniform in $\log a$
ranging from $a \sim 10$~R$_\odot$ up to $10^6$~R$_\odot$.  The
initial eccentricity-distribution is independent of the other orbital
parameters, and is $\Xi(e) = 2 e$.  The mass of the secondary star
$M_2$ is selected with equal probability per mass interval between a
minimum of 0.1~\msun and the mass of the companion star.  We limit
our study to the binaries which do not experience Roche-lobe contact
during their evolution.

We computed models without a kick, with a fixed kick velocity, with a
Maxwellian velocity distribution with a three-dimensional dispersion
of 450~\kms\ (which is close to the distribution proposed by Lyne and
Lorimer 1995, but with a less pronounced low-velocity tail) 
and with the distribution proposed by Hansen \& Phinney
(1996, see also Hartman 1997), which is currently most
favored. Note that the revised velocity distribution from Hansen \&
Phinney (1997) has a smaller contribution from low velocities.
Kicks are, when implied, applied in a random direction.  Only
a binary that survives the first supernova and of which the optical
star is still visible can be identified as a radio pulsar with a
stellar companion.  Whether or not the binary is still detectable
depends on a number of details: the time left before the second
supernova, the age of the pulsar and the mass of its companion.

For instance, a primary with a mass of 12~\msun\ has a stellar
lifetime of $\sim 20$~Myr.  If the companion has a mass of 10~\msun,
and therefore has a stellar lifetime of $\sim 28$~Myr, then the pulsar
is observable with a 10~\msun\ companion for 8~Myr. If the companion
mass was much smaller, say 3~\msun, the pulsar will die long before the
companion burns up.

In the model computations the observable lifetime and the initial
conditions result in the probability distribution for a range of
secondary masses against the age of the pulsar.  The model
computations reveal (see fig.~1) that the most likely companion of a
young pulsar is a 7~\msun or 8~\msun star. This probability drops for
lower mass companions; a binary with a small companion mass is more
likely to be dissociated in the supernova event.  Also for higher mass
companions the probability drops, but now it is mainly the initial
mass function in combination with the mass-ratio distribution which
causes this effect. The effect of the shorter lifetime of the
companion star is observable at the high-mass end of fig.~1; an older
pulsar is less likely to have a massive companion.

\begin{figure}
\vspace*{-2.5cm}
\centerline{\psfig{file=nsnsw_nokick.ps,height=8cm}}
\vspace*{-1.5cm}
\hspace*{-3.25cm}\centerline{\psfig{file=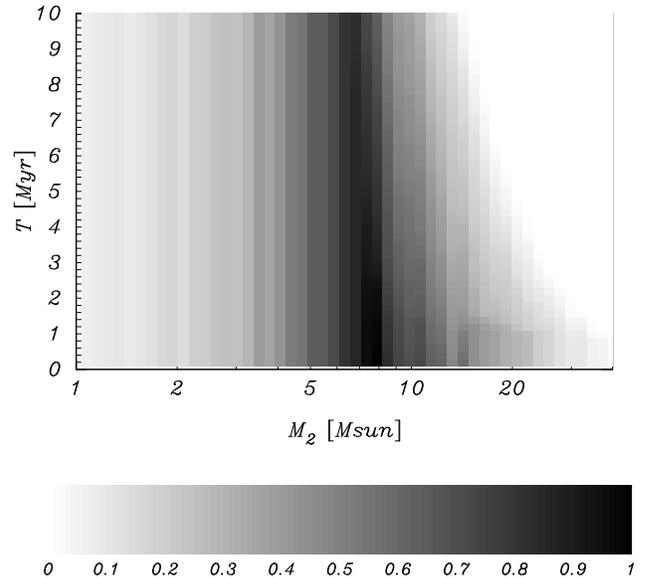,height=8cm}}
\vspace*{-4.5cm}
\caption[]{The probability distribution for secondary mass $M_2$
(X-axis) of a binary as a function of the age of its pulsar
companion (Y-axis) for the model without a kick.  
Darker shades indicate a higher
probability (see lower panel for the scaling).
The distributions for the models with kicks
have similar shapes, except that the total probabilities decrease.
}
\label{fig:binprob}
\end{figure}

\section{Catalog comparison}
For each known radio pulsar with a well-determined position, we have
looked for a counterpart by correlating the pulsar position with an
optical catalog.  The known pulsar population is biased due to various
selection effects, such as the predominance of sources discovered by
Arecibo.  But there is no selection effect against finding pulsars in
wide binaries because they behave exactly as if they were isolated.

The recently published pulsar catalog contains 706 pulsars
(Taylor et al.\ 1993, 1995).  However not all of these pulsars can be
considered in our study. All pulsars which are
member of a globular cluster are excluded; the
evolutionary histories of binaries in globular clusters is very
different from those in the galactic disk.  Also the pulsars which are
known to be the member of an interacting binary system and those which
lie in external galaxies are excluded from consideration, such as
PSR~B1259--63 which has experienced a phase of mass transfer. From
the remaining sample, we rejected the pulsars with positional
accuracies worse than 1 arcsecond and PSR~J0633+1746 (Geminga) of
which the distance is not known.
The distance to each of the remaining 307 pulsars used for
the final analysis is derived
from the dispersion measure using the Taylor \& Cordes (1993) model for the
electron density in the Galaxy and the age obtained from the spin-down
rate (Taylor et al.\ 1995).

The positions from the remaining pulsars are correlated with the Guide
Star Catalog, containing 19 million objects of
which more than 15 million are stars. This is the largest optical
catalog available but has no distance information and the position
accuracy is a few tenth of an arcsecond. The completeness limit of this
catalog is not uniform over the whole sky but it is lower in more
crowded fields.  We estimate the local completeness limit by counting
the number of stars as a function of their magnitude (within bins of
half a magnitude) within a circle of one degree radius around the
position for each pulsar.  The local limit was estimated to be half a
magnitude less than the magnitude bin with the maximum number of
stars.

After completing the correlation, each pulsar is assigned its nearest
optical neighbor from the Guide Star Catalog with a given angular
separation $\delta_{\rm min}$.  This angle is converted into a
lower-limit for the true separation in parsec between the pulsar and
the optical counterpart $d_{\rm psr}$, using the estimated distance to
the pulsar $r_{\rm psr}$.  A separation larger than 0.1~pc
between the pulsar and its optical ``counterpart'' indicates that the
pulsar and the star cannot be associated.  Binaries with major axes
larger than $\sim 0.1$\, pc will be dissociated easily by a
hard encounter of a neighboring star or passing giant molecular clouds
(Binney \& Tremaine 1987). Note that this separation is rather
uncertain.

\begin{table}
\caption[]{
The radio pulsars which have an optical counterpart in the Guide Star
Catalog with a separation (assuming that both objects are at the distance of
the pulsar) smaller than 0.1~pc. The first column lists the pulsars
followed by the derived distance (in kpc), the angular distance of the
optical counterpart (in seconds of arc) and the minimum separation
$d_{\rm min}$ (in parsec).
The last column gives comments
about the pulsars' counterpart, whether it is a single star, a multiple
star or a single non-stellar object.
}
\begin{flushleft}
\begin{tabular}{r|lrll}
PSR  & $r_{\rm psr}$&$\delta_{\rm min}$& $d_{\rm min}$ &
				Comment\\ \hline 
 	  &\unit{kpc}& ['']   & [pc] & \\ \hline	 %
B0454+55 & 0.79     &  15.42  & 0.059 &multiple star \\ 
B0819+74 & 0.31     &  26.12  & 0.039 &multiple star \\	
B0950+08 & 0.12     & 116.60  & 0.068 &single star \\		
B1650--38& 5.12     &   2.12  & 0.053 &single star \\		
B1822--09& 1.01     &  12.70  & 0.062 &single non-star \\	
B1839+09 & 2.49     &   1.82  & 0.022 &single non-star \\	
B1929+10 & 0.17     &   4.87  & 0.004 &single star \\		
B1951+32 & 2.5      &   8.16  & 0.099 &single non-star \\ \hline 
\end{tabular}
\end{flushleft}
\label{tab:pulsars}\end{table}

Table~\ref{tab:pulsars} gives a list of the pulsars wich are
associated with an optical counterpart with a minimum separation of
$<0.1$\,pc.  The pulsars which have a non-stellar object as an optical
counterpart are not considered in our analysis.  Those with correlate
with a multiple star may be of interest but we neglect them here
because our model computations do not incorporate triples.  Pulsar
PSR~B0950+08 is most likely not associated with its single star
counterpart.  It is statistically not unlikely to find a counterpart
at such an angular distance $\delta_{\rm min}$.  
The visual magnitude of the optical counterpart of PSR~B1650--38 is
$11^m.6$ and at the distance of the pulsar (5.12~kpc) a $\sim 15$\msun\
star could be hidden without being noticed.
At the age of the pulsar of 1.7~Myr it is rather unlikely to have such
a massive companion (see fig.~\ref{fig:binprob}). 
The pulsars PSR~B1929+10 has an optical counterpart 
with a magnitude of 12.9 with a minimum separation of 0.004~pc and is
therefore a good binary candidate. The high proper motion of
PSR~B1929+10 of about 100~mas/yr (see Taylor et al. 1995) makes it
possible to test its counterpart for binarity.

\begin{figure}
\vspace*{-3.0cm}
\centerline{\psfig{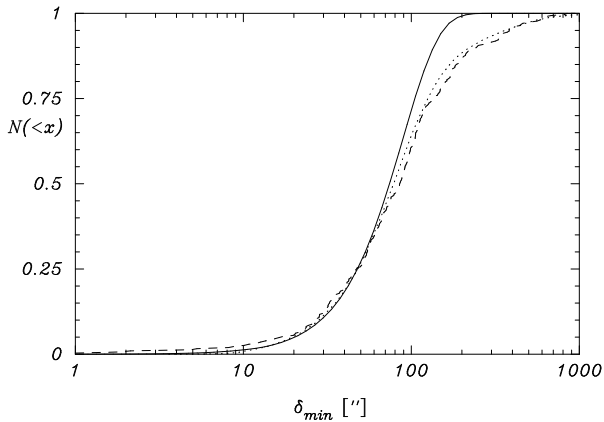}}
\caption[]{Cumulative distribution of the closest distance 
between a pulsar and its nearest star in arcseconds.  The dashed line
gives the results for the correlation between 307 pulsars with the
Guide Star Catalog.

The solid line is the theoretically expected distribution for the
smallest angle between nearest neighbors a set of 19 million 
isotropically distributed objects on the sky.

The dotted line gives the cumulative nearest neighbor distribution of
about 30000 randomly selected single stars in the Guide Star Catalog.
A Kolmogorov-Smirnov test reveals that the nearest neighbor
distribution for radio pulsars (dashed) and the randomly selected
entries in the Guide Star Catalog (dots) are not distributed differently
with a confidence level of 98.7\%.
}
\label{histo_distance}
\end{figure}

Fig.~\ref{histo_distance} gives the cumulative distribution of angles
between 307 radio pulsars and their nearest entry in the Guide Star
Catalog.  The expected distribution (solid line) of smallest angles
assumes that stars and pulsars are distributed isotropically on the
sky. The deviation of the observed distributions indicates that
pulsars and stars are not distributed isotropically; radio pulsars and
single stars are confined to low Galactic latitude.  

\section{Comparison of the model results with the observations}
Comparison of the pulsar catalog with the Guide Star Catalog provides the
following information for each pulsar: The distance to the sun
(determined from the dispersion measure), the age (derived from the
pulse period and its derivative), the magnitude limit of the Guide
Star Catalog in the direction of the pulsar and the angular distance
between the pulsar and the nearest neighboring star $\delta_{\rm
min}$.

The distance to each radio pulsar $r_{\rm psr}$ together with the
local magnitude limit of the Guide Star Catalog in the direction of
the pulsar is used to derive the maximum absolute magnitude for a
companion of the pulsar which should have been noticed.  The
transformation from this relative magnitude limit $m_v$ to the
absolute magnitude $M_v$ is performed using the following relation:
$m_v - M_v = 5 \log r_{\rm psr} -5 + A r_{\rm psr}$ where we used $A =
1.6$~mag\,kpc$^{-1}$ to correct for interstellar extinction and
$r_{\rm psr}$ in kpc.  (Note that an extinction of $A =
1.6$~mag\,kpc$^{-1}$ is rather high for objects which have a large
scale height above the galactic plane, but we are interested in lower
limits to the observability of companion stars.)

Using the absolute magnitude limit for each pulsar together with
a mass-luminosity relation for zero-age
main-sequence stars provides a 
{\em minimum} mass of a main-sequence star $m_{\rm min}$
which could have been seen at the distance of this pulsar.

We compute the probability $P_b$ that a pulsar has retained its
companion and that this star is still visible.
This $P_b$ is obtained by integrating in fig.~1 from $m_{\rm
min}$ to the maximum companion mass at the age of the pulsar with a
binwidth of 1~Myr.  

The model computations provide us with the number of single pulsars
from dissociated binaries $N_s$ and pulsars which are still member of a
binary after the first supernova $N_b$. For $N_b$ we only selected
those pulsars with a major axis large enough that no Roche-lobe
overflow occurred and small enough that the binary is not dissociated
by a close encounter in the Galactic plane,
i.e.\ between $\sim 10^4$~\rsun\ and 0.1~pc. 
Binaries with smaller and larger major axes fall beyond the scope of
this discussion.
For simplicity we assume 100\% binarity among the progenitors 
but we also perform computations with 50\% binarity (see
Tab.~\ref{tab:table}).  
For each individual pulsar we can now 
compute the expectation value $E_{\rm obs}$ that it has an observable
companion by correcting $P_b$ for the fraction of pulsars in binaries:
\begin{equation}
	 E_{\rm obs} = P_b \frac{N_b}
	     	                {N_s + N_b}.
\end{equation}

The total expected number of optical counterparts among radio pulsars
which should have an entry in the Guide Star Catalog is given by:
\begin{equation}
	E = \sum_i^{N_{\rm psr}} (E_{\rm obs})_i,
\end{equation}
where the sum goes over all selected radio pulsars $N_{\rm psr} = 307$
in the catalog.  A value of $E$ larger than unity indicates that we
expect to see at least one binary among the observed radio pulsars.
Table~\ref{tab:table} gives this probability for the various models.
The pulsars distances are uncertain with a factor of two or
so, but our results are not very sensitive to the distance scaling.
If the distances to the pulsars are increased 
with a factor of two low mass companions become hard to observe but
the probability for having such a companion is small anyway (see
fig.~\ref{fig:binprob}).  

\begin{table}
\caption[]{
The probabilities for the set of observed pulsars to have a
companion which should have an entry in the Guide Star Catalog. The first row
gives the value of the fixed velocity kick
in km/s and the distribution proposed
by Hartman (1997, see also Hansen \& Phinney 1996) for the ninth
column ($f_H$) and a 
Maxwellian ($f_M$) with a dispersion of 450~\kms\ for the last.
The second and third row give the expectation values $E$ 
for the computations with 100\% and $\sim 50\%$ binarity,
respectively, for the total number of expected binaries found.
}
\begin{flushleft}
\begin{tabular}{l|ccccccc|cc} \hline
$v_k$& 0 & 1 & 5 & 10 & 15 & 20 & 25 & $f_H$ & $f_M$ \\\hline
100\%& 52& 40& 11& 3.2 &1.2&0.45&0.17&0.32&0.01\\ 
50\% & 26& 20& 5.6& 1.6 &0.59&0.23&0.09&0.17&0.00\\ \hline
\end{tabular}
\end{flushleft}
\label{tab:table}\end{table}

\vspace*{-0.25cm}
\section{Conclusion}

Comparison of the positions of 307 known radio pulsars on the sky with
those of visible stars reveals that a tiny fraction of the radio
pulsars can be associated with an entry in the Guide Star Catalog.
The associated fraction is considerably smaller than what is expected
on statistical grounds if there are no kicks.  An asymmetry in all
supernov\ae\ of at least 10~km/s
satisfactorily explains this underabundance of optical counterparts.
This is in agreement with the results of Cordes \& Chernoff (1997) who
find no evidence for a low velocity tail.
The kick velocity distribution proposed by Hansen \& Phinney (1996)
and Hartman (1997) is also sufficient to explain this lack.
However, if the possible counterparts   
found in this paper are indeed
associated with a radio pulsar the kick-velocity distribution
requires a contribution from velocities below $\sim 10~$\kms\ and
provides evidence for the presence of a low-velocity tail.

If this kick-velocity
distribution represents indeed the intrinsic velocity kick received by
neutron stars, the observed number of single radio pulsar should be 
at least a factor 3 larger than at present before one can hope to find
a very wide binary which contains a radio pulsar.

The only pulsar which is possibly a member of a very wide binary is
PSR~B1929+10. Its companion could be a 2~\msun\ main-sequence star at the
distance of the pulsar. 

\acknowledgements
We would like to thank Dipankar Bhattacharya and Edward van den
Heuvel for stimulating discussions and comments on the manuscript.
The referee Brad Hansen is acknowledged for his detailed comments.
This investigation is supported by NWO under Pioneer grand PGS~78-277
to F.~Verbunt and by Spinoza grant 08-0 to E.~P.~J.~van den Heuvel.

\end{document}